\documentclass[final]{elsarticle}

\usepackage[utf8]{inputenc} 
\usepackage[T1]{fontenc}    
\usepackage{hyperref}       
\usepackage{url}            
\usepackage{booktabs}       
\usepackage{amsfonts}       
\usepackage{nicefrac}       
\usepackage{microtype}      
\usepackage{xcolor}         

\usepackage{overpic}
\usepackage{amsmath}
\usepackage{amssymb}
\usepackage{utfsym}
\usepackage{bbm}
\usepackage{threeparttable}
\usepackage{geometry}
\begin{document}

\begin{frontmatter}

\title{3D Matting: A Soft Segmentation Method Applied in Computed Tomography}


\author[1st_address,2nd_address,3th_address]{Lin Wang}
\author[1st_address]{Xiufen Ye\corref{mycorrespondingauthor}}
\author[2nd_address]{Donghao Zhang}
\author[3th_address]{Wanji He}
\author[2nd_address,3th_address]{Lie Ju}
\author[3th_address]{Xin Wang}
\author[2nd_address,3th_address]{Wei Feng}
\author[3th_address]{Kaimin Song}
\author[3th_address]{Xin Zhao}
\author[2nd_address,3th_address]{Zongyuan Ge\corref{mycorrespondingauthor}}
\cortext[mycorrespondingauthor]{Corresponding author\\ 
\textit{
\indent\; Email address: zongyuan.ge@monash.edu (Zongyuan Ge) \\ 
\indent\qquad\qquad\qquad\quad\ yexiufen@hrbeu.edu.cn (Xiufen Ye)}}

\address[1st_address]{College of Intelligent Systems Science and Engineering, Harbin Engineering University, Harbin, China}
\address[2nd_address]{Monash Medical AI Group, Monash University, Clayton, Australia}
\address[3th_address]{Beijing Airdoc Technology Co., Ltd., Beijing, China}

\begin{abstract}
    Three-dimensional (3D) images, such as CT, MRI, and PET, are common in medical imaging applications and important in clinical diagnosis. 
    Semantic ambiguity is a typical feature of many medical image labels. It can be caused by many factors, such as the imaging properties, pathological anatomy, and the weak representation of the binary masks, which brings challenges to accurate 3D segmentation.
	In 2D medical images, using soft masks instead of binary masks generated by image matting to characterize lesions can provide rich semantic information, describe the structural characteristics of lesions more comprehensively, and thus benefit the subsequent diagnoses and analyses.
	In this work, we introduce image matting into the 3D scenes to describe the lesions in 3D medical images.
	The study of image matting in 3D modality is limited, and there is no high-quality annotated dataset related to 3D matting, therefore slowing down the development of data-driven deep-learning-based methods. 
	To address this issue, we constructed the first 3D medical matting dataset and convincingly verified the validity of the dataset through quality control and downstream experiments in lung nodules classification.
	We then adapt the four selected state-of-the-art 2D image matting algorithms to 3D scenes and further customize the methods for CT images.
    Also, we propose the first end-to-end deep 3D matting network and implement a solid 3D medical image matting benchmark, which will be released to encourage further research\footnote{Url for codes and dataset: \url{https://github.com/wangsssky/3DMatting}.}.
\end{abstract}

\begin{keyword}
3D Matting\sep Pulmonary nodules\sep Soft Segmentation\sep Thoracic CT\sep Uncertainty 
\end{keyword}

\end{frontmatter}

\section{Introduction}

\begin{figure}[!t]
	\begin{center}
		\begin{overpic}[width=\textwidth]{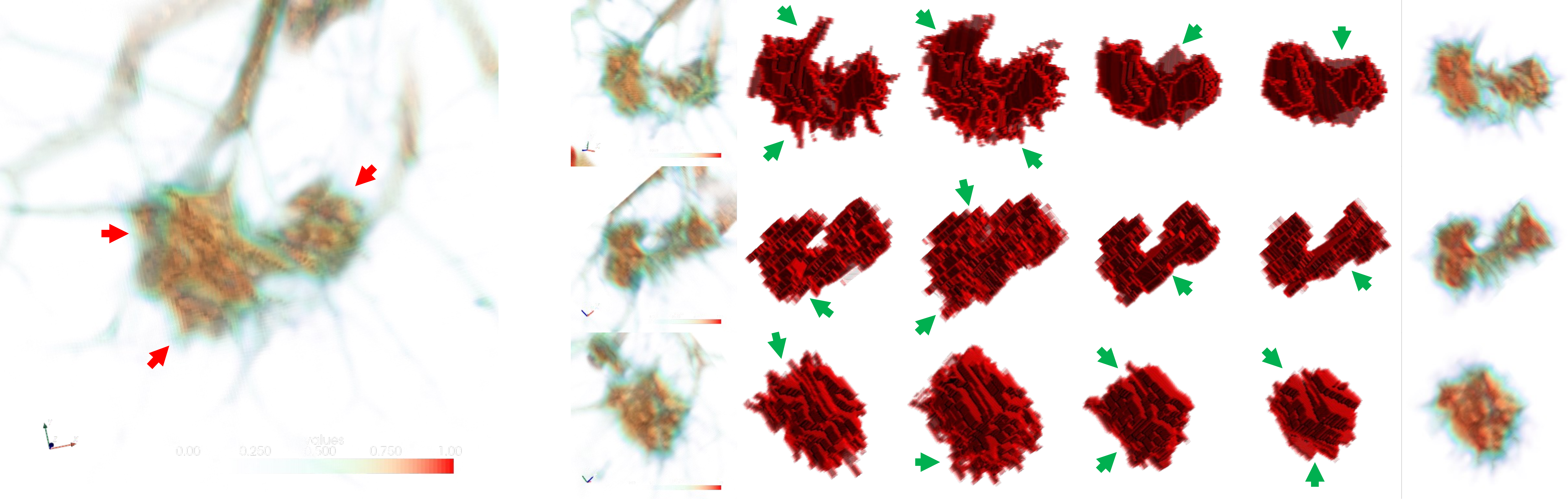}
			\put(-2,15){A}
			\put(33,25){B}
			\put(33,15){C}	
			\put(33,5){D}
			\put(41,-2.5){(a)}
			\put(51.5,-2.5){(b)}
			\put(62,-2.5){(c)}
			\put(72.5,-2.5){(d)}
			\put(83,-2.5){(e)}
			\put(93.5,-2.5){(f)}
		\end{overpic}
	\end{center}
	\caption{
		An example of describing lung nodules~\cite{armato2004lung} with binary masks and alpha mattes. 
		We show a pulmonary nodule sample (A) in multiple views (B-D).
		Because of the fuzziness of nodule boundaries (some examples are indexed by red arrows), it is difficult to keep the binary masks (b-e) marked by different clinicians consistent (indexed by green arrows).
		Moreover, it is challenging to annotate the edges of fuzzy lesions accurately and maintain the annotation continuity between continual slices in the 3D scene.
		However, the ground truth soft masks (f), i.e., the alpha mattes, have a better ability to represent the details of the lesions, with better continuity in between the slices.
        The redder the color, the more likely it is to be part of a lesion.
        }
	\label{fig:fuzziness}
\end{figure}
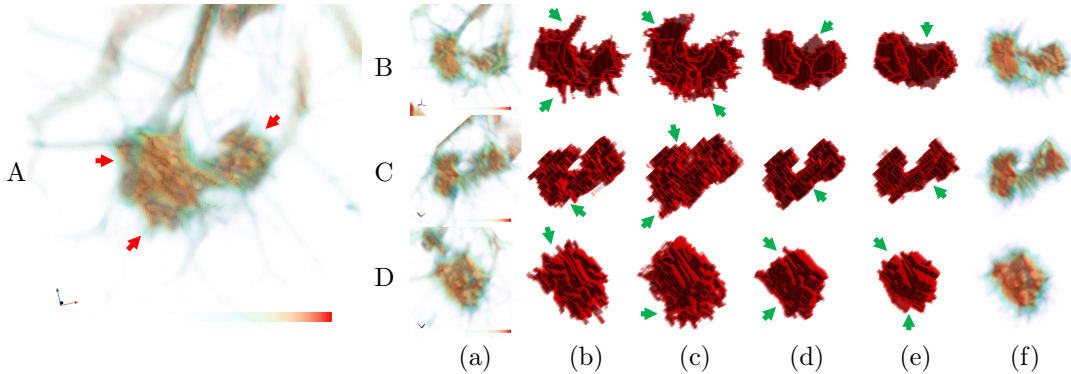

Due to image noises, the occlusion of human tissues, the principles of medical imaging, and the anatomical structure characteristics of lesions, fuzzy boundaries are inevitable and ubiquitous in medical images~\cite{kohl2019hierarchical,wang2021medical}.
In segmentation tasks, binary masks are the most commonly used to identify lesion regions. However, the fuzzy boundaries impede the accurate recognition of the lesion regions, which results in the uncertainty in the annotation procedure, adversely affecting the downstream diagnoses and treatments. Figure~\ref{fig:fuzziness} gives an example of fuzziness in medical images.

Many studies have been proposed to mitigate the effects of ambiguity in medical image segmentation. Multi-annotated datasets are proposed to reduce the labelling bias~\cite{armato2004lung,codella2018skin}, and probabilistic models~\cite{baumgartner2019phiseg,kohl2018probabilistic,kohl2019hierarchical} attempt to describe the lesion distribution, etc. 
However, due to the complex causes of ambiguity, it is still challenging to eliminate it.

Instead of identifying perfect lesion boundaries that are difficult to annotate using binary-style masks, medical matting~\cite{wang2021medical} makes use of the ambiguous regions. 
It introduces image matting into the medical scenes, and regards medical images as a mixture of lesions and healthy tissues.
This mixing factor is known as \textit{alpha matte}.
In such a way, the fuzzy boundaries are taken as the transition area from pure lesions to pure healthy tissues. 
The alpha matte can be used as a soft mask to describe the anatomical structures of lesions more comprehensively than a binary mask, especially in the fuzzy area.

In addition, the fuzzy areas contain important diagnostic information.
For example, the fuzzy area around lung nodules in the lung CT images may refer to two kinds of the border, the indistinct border and amorphous ground-glass shadow, which is vital for clinical staging of nodules~\cite{dyer2020implications}.
Therefore, by keeping more details, medical matting instead of binary segmentation provides more possibilities for down streaming tasks, such as nodule grading and precise radiotherapy.

The previous works mainly focused on 2D medical images~\cite{wang2021medical,zhao2020improving,fan2018hierarchical}, and the existing research on 3D medical image matting is very limited in quantity and methodology. 
To the best of our knowledge, only~\cite{zhong20173d,zhong2018improving,liu2020three,khan2022deep} have touched upon the problem of the 3D matting, and they are mainly based on 2D \textit{closed-form} matting~\cite{levin2007closed}. There is no deep-learning-based method investigated.
To address this, we adapt the matting methods to 3D medical image scenes, including four traditional methods and a deep-learning-based method, as a more accurate approach for lesion segmentation and description, especially for the fuzzy areas.

As we know, this is the first work to explore the possibility of matting to solve fuzzy boundary problems in the scenario of 3D medical image segmentation. Furthermore, it is also the first attempt to deploy deep-learning-based matting, rather than the traditional matting method, to 3D image data to achieve automatic inference without human intervention.
\textbf{Firstly}, due to the lack of available datasets in the 3D matting scenario, we have created a publicly accessible dataset based on the LIDC-IDRI, hoping to benefit the 3D matting research community.
Using this dataset, it is verified quantitatively that the alpha matte contains more diagnostic information than the binary mask.
\textbf{Furthermore}, four state-of-the-art traditional 2D matting methods are adapted to 3D scenarios and further customized to CT images, making the dataset built in a semi-automatic approach.
\textbf{Finally}, the 3D medical matting network, a benchmark 3D deep-learning-based model, is proposed as an end-to-end automatic matting network for pulmonary nodules.

\section{Related Work}
There have been some attempts at soft segmentation in the medical field. For example, Kats et al. ~\cite{kats2019soft} claim that pixels around lesions also have diagnostic information and assign them with a soft label in training to improve the segmentation performance. Dai et al. ~\cite{dai2022soft} use soft masks in data augmentation, which mixups the lesion with the image by using a soft coefficient at the boundaries of the lesions. However, such soft masks do not reflect the structural information of the lesions.

Image matting~\cite{aksoy2017designing,cai2019disentangled,chen2013knn,chuang2001bayesian,forte2020fbamatting,levin2007closed,lutz2018alphagan,wang2008image,xu2017deep} uses the mixing coefficient $\boldsymbol{\alpha}$, also known as the \textit{alpha matte}, to decompose the image $\mathcal{I}$ to foreground $\mathcal{F}$ and background $\mathcal{B}$, or lesions and its surrounding tissues in medical images~\cite{wang2021medical}, which can be defined as
\begin{equation}
	\mathcal{I}_i = \boldsymbol{\alpha}_i \mathcal{F}_i + (1-\boldsymbol{\alpha}_i) \mathcal{B}_i.
	\label{eq:image_matting}
\end{equation}
Image matting is a particular type of image segmentation that uses the alpha matte, a soft mask, to describe the target. It is beneficial for image/video editing when dealing with the blur boundaries. In medical applications, alpha matte can better substitute for binary masks to depict lesions with more details preserved~\cite{wang2021medical}. 

\begin{figure}[!t]
	\begin{center}
		\begin{overpic}[width=\textwidth]{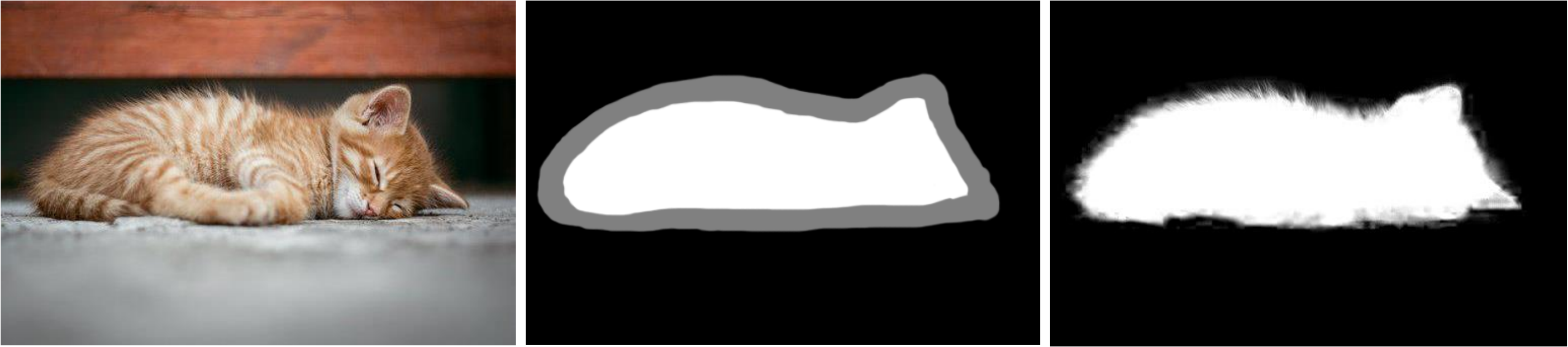}
			\put(10,-2.5){(a) input image}
			\put(45,-2.5){(b) trimap}
			\put(76,-2.5){(c) alpha matte}			
		\end{overpic}
	\end{center}
	\caption{
		A example of 2D matting. The traditional matting method infers alpha matte (c) with the help of the prior information provided by trimap (b).
    }
	\label{fig:trimap}
\end{figure}

However, since only $\mathcal{I}$ is known in the four variables in Eq.~\ref{eq:image_matting}, matting is an ill-posed problem~\cite{yao2017comprehensive}, which is challenging to solve directly.
A common practice is to reduce the problem complexity by introducing a prior map called \textit{trimap} as constraints, indexing the regions of the foreground $\mathcal{R}_f$, background $\mathcal{R}_b$, and unknown $\mathcal{R}_u$. Figure~\ref{fig:trimap} gives an example of trimap and alpha matte of 2D natural image for intuitive understanding.
Therefore, according to whether the trimap is used, the matting methods can be divided into trimap-based and trimap-free methods. The trimap-based methods generally achieve better performance as the trimap provides additional information. However, generating the trimap requires extra manual labor and limits the application in practice~\cite{chen2022pp}.

According to whether deep-learning techniques are used, image matting methods can be divided into traditional methods~\cite{aksoy2017designing,chen2013knn,chuang2001bayesian,levin2007closed}  and deep-learning-based methods~\cite{cai2019disentangled,forte2020fbamatting,lutz2018alphagan,xu2017deep}. 
Thanks to the latest development in artificial intelligence, the deep-learning-based methods, driven by extensive data, tend to have better performance and a more comprehensive range of applications than traditional methods. 
However, the traditional methods are still popular because they require no tedious training phase and have better generalization in new applications. 

Matting in 3D is not well studied as the majority of the natural images are 2D images. However, because a large proportion of the images of various modalities in medical scenarios are in 3D, there is a strong demand for 3D matting in medical applications. 
The research on 3D medical matting is limited. All of them are based on traditional methods and act as an auxiliary means of binary segmentation. 
For example, Zhong~\cite{zhong20173d,zhong2018improving} adapts \textit{closed-form} matting~\cite{levin2007closed} to 3D, and uses alpha mattes as probability maps of tumors in calculating the region cost for PET-CT co-segmentation.
Liu~\cite{liu2020three} also uses a 3D \textit{closed-form} matting for organ model extraction for Virtual Human Project with significant efficiency improvement.
These studies show the potential of matting in 3D scenes, but they are not comprehensive and in-depth enough compared to the critical importance of 3D data in medical diagnosis.
This work investigates the traditional matting methods in 3D and provides a trimap-free deep-learning-based solution.

\section{Benchmark Dataset}

Although the deep-learning-based methods~\cite{cai2019disentangled,forte2020fbamatting,lutz2018alphagan,xu2017deep} are the frontier of 2D matting research nowadays and generally achieve better performance, they require a relatively large amount of samples for training. 
However, unfortunately, because the matting of 3D scenes has not been well studied, there is no available dataset in the 3D medical matting research community. 
The lack of datasets severely limits the research of 3D medical matting.
In order to study the 3D matting quantitatively and facilitate the research community, we build a 3D matting dataset of lung nodule CT images based on the LIDC-IDRI dataset~\cite{armato2004lung}.

\paragraph{Challenges of Manual Labelling}
Manual labelling is slow and human-labor intensive, which is especially true for 3D data. 
Generally, it takes an experienced radiologist at least 15 minutes to label a 2D image, which is hardly affordable to label a 3D medical image with thousands of slices. 
Moreover, the slice-by-slice labelling may lead to the discontinuity between neighboring slices.
Therefore, we adopt the method of alpha matte labelling in 2D images~\cite{2016Deep}, combined with manual screening, to achieve semi-automation of annotation, improve annotation efficiency, and ensure the annotation quality.

\paragraph{Labelling Pipeline and Quality Control}
For 2D portrait matting, Shen~et al.~\cite{2016Deep} deploy the \textit{closed-form} matting~\cite{levin2007closed} and KNN matting~\cite{chen2013knn} to generate alpha mattes, i.e., the soft segmentation labels, and then select the better one to describe the portrait as the ground truth (GT). 
To set up the research for our 3D matting task, four state-of-the-art traditional 2D matting methods, which are adapted into 3D scenes and optimized for CT images (introduced in Section~\ref{traditional_methods}), 
are employed to generate a set of alpha mattes for each 3D nodule image. Then we manually chose the one which depicts the nodule the best 
as the GT label (See Figure~\ref{fig:fuzziness}~(f) as an example.). However, if all the generated alpha mattes cannot meet our quality requirement, the corresponding nodule image will be discarded. 

\begin{figure}[!t]
	\begin{center}
		\includegraphics[width=\textwidth]{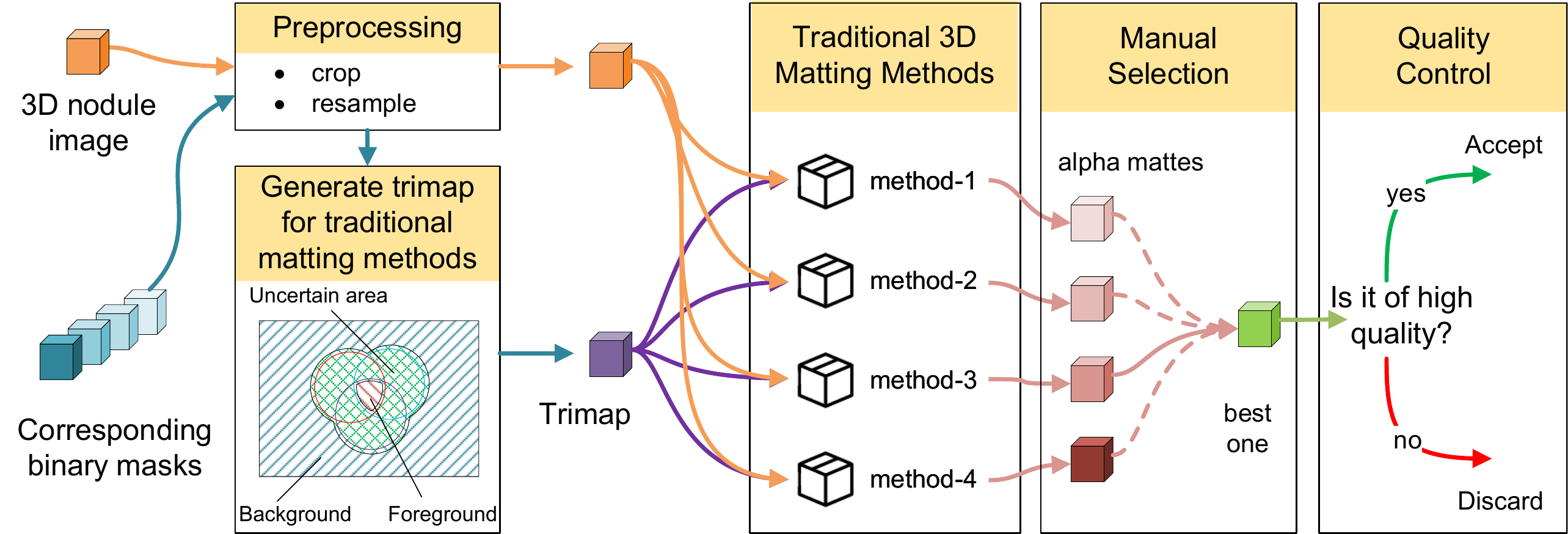}
	\end{center}
	\caption{
		The schematic diagram of labelling pipeline.
		The dataset is generated in a semi-automatic way, using the customized traditional 3D matting methods and the multiple binary annotations. The quality control keeps the dataset to a high-quality standard.
	}
	\label{fig:labelling_pipeline}
\end{figure}

Specifically, for preprocessing, each nodule is centered and cropped transversely with the size of $128\!\times\!128$ and sagittally with 3-slice padding at both ends. 
Then the spacing of these 3D patches is resampled to be $0.5 mm$ in every dimension. 
The traditional methods used to create the alternative alpha mattes do not require training but only need to provide the definite foreground, background, and uncertain areas as constraints (i.e., trimap). In the 2D scenario, such auxiliary maps are annotated manually. In this work, we utilize the information of the multi-annotated binary masks to reduce the workload further.
Since each nodule is labelled by four binary masks, 
we take the intersection of the four binary masks as the foreground, the complement of the union as the background, and the remaining as the unknown region.
Practically, we dilate the unknown region to reduce the bias of manual labelling.
Finally, after manual quality control filtering, we obtain the 3D medical matting dataset, which includes 864 3D nodule images from 542 patients.
Figure~\ref{fig:labelling_pipeline} illustrates the pipeline of the annotation.
Experiments in Section~\ref{Exp_Trad_3D_Matting} illustrate the validity of the dataset.


\section{Traditional Matting Method Adaptation to 3D Scenarios}\label{traditional_methods}

\subsection{3D Traditional Matting Methods}

According to the online matting benchmark\footnote{Url for the online matting benchmark: \color{magenta}{\url{http://alphamatting.com}}.}, four state-of-the-art matting methods, i.e., the \textit{closed-form} matting~\cite{levin2007closed} (CF), \textit{KNN} matting~\cite{chen2013knn} (KNN), \textit{learning-based} matting~\cite{zheng2009learning} (LB), and \textit{information-flow} matting~\cite{aksoy2017designing} (IF), have been selected as the basis of our research because they rank first among the traditional matting methods.
According to the taxonomy in~\cite{yao2017comprehensive}, CF and LB are affinity-based methods, which construct the relationship between neighboring pixels (e.g., pixels defined in a local window) first and then propagates the label information from the foreground $\mathcal{R}_f$ and background $\mathcal{R}_b$ to the unknown regions $\mathcal{R}_u$. KNN and IF are hybrid methods that use both information from local pixels and non-local pixels. 

Following these traditional 2D solutions, we expand the space from 2D to 3D. 
In general, the solution of alpha matte in these methods can be written as the function:
\begin{equation}
	\boldsymbol{\alpha} = f\left(\mathcal{I}, \mathcal{S}, \mathcal{D},\lambda\right).
\end{equation}
$\mathcal{I}$ is the input 3D image containing $N$ voxels. 
$\mathcal{S}$ is a $N\!\times\!1$ matrix, providing the prior restriction information $\mathcal{R}_f$ and $\mathcal{R}_b$. $\mathcal{D}$ is a $N\!\times\!N$ diagonal matrix indexing the position of the restriction, and $\lambda$ is the weight of the restriction information.
In the case of 3D CF, the alpha matte is computed as:

\begin{equation}
	\begin{aligned}
		\boldsymbol{\alpha}&=\mathit{arg}\, \underset{\boldsymbol{\alpha}}{min}\ 
		\boldsymbol{\alpha}^T \mathcal{L} \boldsymbol{\alpha}+\lambda(\boldsymbol{\alpha}-\mathcal{S})^T\mathcal{D}(\boldsymbol{\alpha}-\mathcal{S}),\\
		\mathcal{L}(p,q)&=\sum_{n|(i,j) \in w_n}
		\left(
			\delta_{i,j}-\frac{1}{k^3}(\frac{1}{\sigma_n^2 + \frac{\epsilon}{k^3}}(\mathcal{I}_{ni}-\mu_n)(\mathcal{I}_{nj}-\mu_n) + 1)
		\right),
	\end{aligned}
\end{equation}
where $\mathcal{L}$ is a $N\!\times\!N$ \textit{matting laplacian matrix}. $\mathcal{L}(p,q)$ is the $(p,q)$-th entry of $\mathcal{L}$, and $(i,j)$ is the corresponding entry of $(p,q)$ in $k\!\times\!k\!\times\!k$ size local window of voxels $w_n$.  
$\mathcal{I}_{n(\cdot)}$ is the $(\cdot)$-th voxel in $w_n$ of $\mathcal{I}$. 
$\delta_{i,j}$ is the Kronecker delta. $\mu_n$ and $\sigma_n^2$ are the mean and variance of $\mathcal{I}$ in $w_n$, while $\epsilon$ is a coefficient of the regularizer. 
The details of the derivation is provided in Appendix.
The solutions of the left methods can refer to the codes. 

\subsection{Optimization for Medical Images}
Unlike the common natural images, the manually indexed foreground in medical images may not be such a pure foreground or of pure lesion tissues. For example, tumors often grow close to or even mix with healthy tissues. In addition, there is crucial structural information in the mixture, and simple homogenization of the foreground area will lose potential diagnostic information. Hence, it is inappropriate to set the alpha matte of the aforementioned foreground in medical images to be $1$. 
To address the problems in transferring 2D matting methods directly to 3D, we modify the methods toward the medical images, and use CT as an example here. 

It is assumed that the areas within a specific Hounsfield Unit\footnote{The Hounsfield Unit in CT is similar to the intensity in the gray image. It generally ranges from -1000 HU to +~2000 HU for human tissues. In practice, a specific window is selected for better focusing on the particular observing target.} (HU) range $(l_{low}, l_{high})$ in the foreground are pure lesions, and the further away from this range, the lower the percentage of lesion tissues. 
Then the foreground of the alpha matte can be written as:
\begin{equation}
	\underset{i\in \mathcal{R}_f}{\boldsymbol{\alpha}_i}= 
	\begin{cases}
	p(\mathcal{I}_i), & l_{high} < \mathcal{I}_i\\
	1, &l_{low} \leq \mathcal{I}_i \leq l_{high} ,\\
	q(\mathcal{I}_i), & \mathcal{I}_i < l_{low}
	\end{cases}
\end{equation}
where $p(\cdot)$ and $q(\cdot)$ are monotonic mappings over a value domain of 0 to 1.
In this way, the alpha matte is associated with the HU, making the alpha matte physically meaningful.
For simplicity but without generality, we set the upper limit of the CT observation window as $l_{low}$ and set $q(\cdot)$ as a linearly increasing function defined over the window width.
By applying the calibrated foreground as a constraint in $\mathcal{S}$, we achieve the optimized 3D traditional matting methods, which denote CF+, KNN+, LB+, and IF+, respectively.

For the sake of description, we refer to the 3D traditional matting methods as \textbf{traditional methods} and their optimized counterparts as \textbf{optimized methods} in the following sections.

\section{3D Medical Matting Network}\label{methods}

\begin{figure}[!t]
	\begin{center}
		\includegraphics[width=\textwidth]{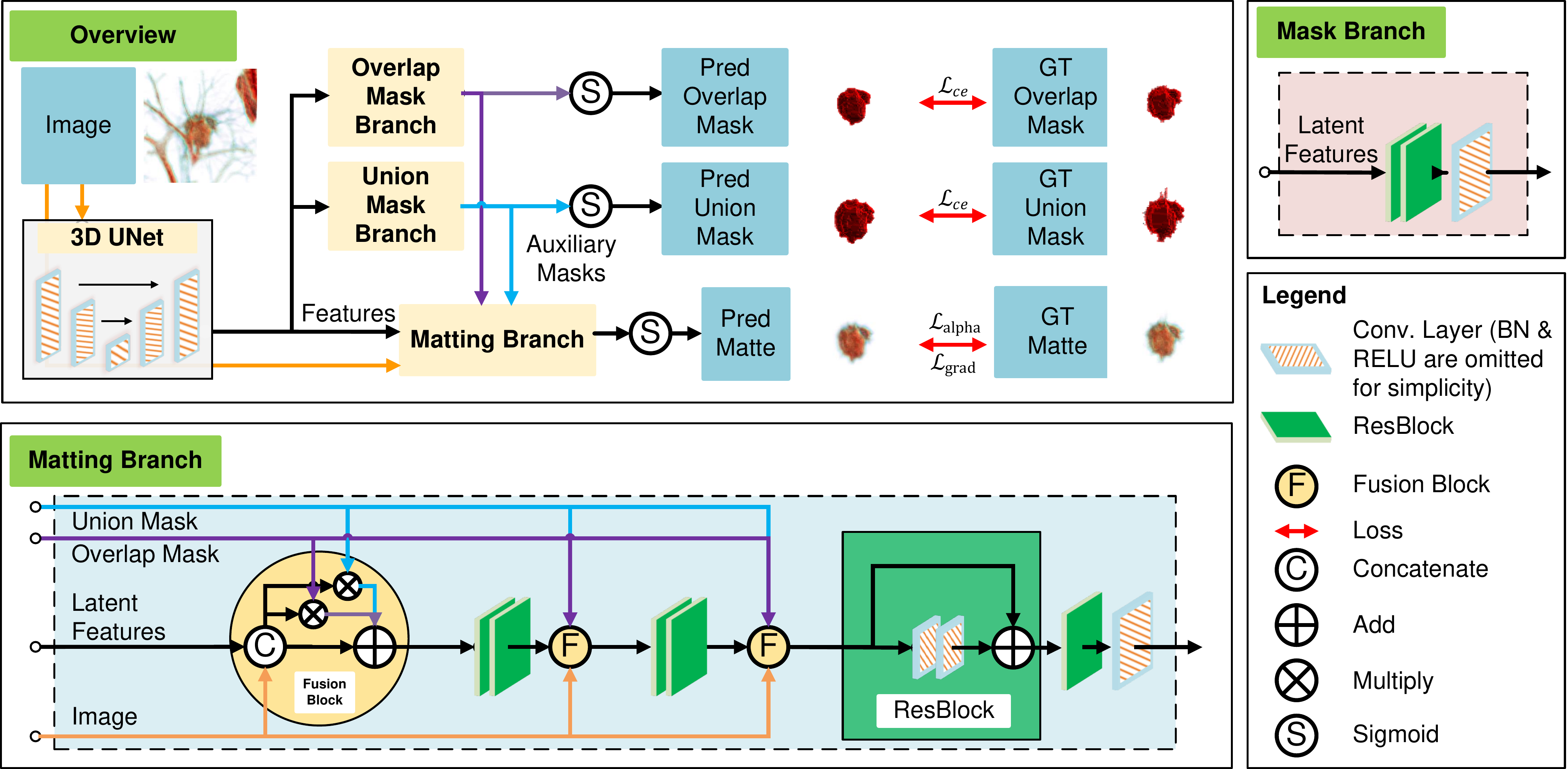}
	\end{center}
	\caption{
		The schematic diagram of the end-to-end 3D Medical Matting Network.
		As a benchmark model, we keep the network elegant. The network contains three parts, i.e., the 3D UNet, the auxiliary mask branches (overlap/union), and the matting branch. A vanilla 3D UNet is used as the feature extractor. The overlap/union mask branch learns the overlap/union of the manual labelled binary masks as auxiliary masks. The matting branch takes these auxiliary masks as guidance and outputs the predicted alpha matte. In this way, there is no need for the manual input of a priori information, and the process is fully automatic. 
	}
	\label{fig: structure}
\end{figure}

The methods proposed in Section~\ref{traditional_methods} do not need model training. 
However, their inferences require human intervention to provide a trimap indexing the foreground and background, which restricts the practical applications.
Thus, we propose an end-to-end deep-learning-based model named \textbf{3D} \textbf{M}edical \textbf{M}atting network (\textbf{3DMM}) as a benchmark of 3D matting, which provides a fully automatic matting inference without human involvement.

As the first deep 3D matting network, 3DMM is focused on feasibility verification rather than pursuing high performance.
Hence, we adopt widely used components referring to the successfully applied 2D matting networks~\cite{cai2019disentangled,2016Deep,xu2017deep} and simplify them.
Overall, the 3DMM, a multi-task network\cite{wang2021medical}, consists of four sub-modules, i.e., the \textbf{3D UNet}, \textbf{Overlap / Union mask} branch, and \textbf{Matting} branch.

\subsection{Network Design}
Firstly, the \textit{3D version of vanilla UNet} is used as the base feature extractor for the subsequent three sub-modules.
Secondly, since the matting task usually requires a priori information of the lesion region as a constraint, two auxiliary networks, the \textit{Overlap mask branch} and the \textit{Union mask branch}, are proposed to learn the overlap and union of the manual binary masks, providing information of the identified core lesion region and the rough contour, respectively.
Besides benefiting the matting performance, these two masks also represent the lesion region in different confidence levels, which expand the applicable scenarios.
Their network structures are identical: a group of Resblocks~\cite{he2016deep} followed by a $1\!\times\!1\!\times\!1$ convolutional layer.
At last, the \textit{Matting branch} consisting of three consecutive groups of Resblocks and a $1\!\times\!1\!\times\!1$ convolutional layer, generates the predicted alpha matte.
Before each group of Resblocks, we fuse the auxiliary information obtained from the two mask branches with the features.

Overall, the 3DMM is a multi-task network that takes a 3D image as input and outputs three predictions, i.e., the union mask, the overlap mask, and the alpha matte.
Figure~\ref{fig: structure} provides a schematic view of the 3DMM framework.

\subsection{Losses}
We use the cross-entropy losses $\mathcal{L}_\mathit{ce_{overlap}}$ and $\mathcal{L}_\mathit{ce_{union}}$ for the overlap mask and the union mask prediction, respectively. 
The absolute difference $\mathcal{L}_\mathit{alpha}$ and the gradient difference $\mathcal{L}_\mathit{grad}$ between the predicted alpha matte $\tilde{\boldsymbol{\alpha}}$ and the GT alpha matte $\boldsymbol{\alpha}$ are deployed to the matting branch~\cite{2016Deep,xu2017deep,chen2022pp}.
The definitions are as follows:
\begin{equation}
	\begin{aligned}
		\mathcal{L}_\mathit{mask}&= \mathcal{L}_\mathit{ce_{overlap}} + \mathcal{L}_\mathit{ce_{union}}, \\
		\mathcal{L}_{alpha}&=\frac{1}{\sum\mathbbm{1}_{\tilde{\boldsymbol{\alpha}}}} \sum\nolimits_{i \in \tilde{\boldsymbol{\alpha}}} 
		\left (1 + \mathbbm{1}_{\mathcal{M}_\mathit{union}}(i) \right ) 
		\left \|\tilde{\boldsymbol{\alpha}}_i-\boldsymbol{\alpha}_i\right \|_1, \\
		\mathcal{L}_{\mathit{grad}}&=\frac{1}{\sum\mathbbm{1}_{\tilde{\boldsymbol{\alpha}}}}\sum\nolimits_{i \in \tilde{\boldsymbol{\alpha}}}
		\left (1 + \mathbbm{1}_{\mathcal{M}_\mathit{union}}(i) \right ) 
		\left \|\nabla\tilde{\boldsymbol{\alpha}}_i-\nabla\boldsymbol{\alpha}_i\right \|_1,
	\end{aligned}
	\label{eq:losses}
\end{equation}
where $\mathcal{M}_\mathit{union}$ is the GT union mask used to amplify the weight of the target regions.
Finally, the total loss $\mathcal{L}_\mathit{total}$ can be written as:
\begin{equation}
	\mathcal{L}_\mathit{total}=\mathcal{L}_\mathit{mask} + \alpha\mathcal{L}_\mathit{alpha} + \beta\mathcal{L}_\mathit{grad},
	\label{eq:total_loss}
\end{equation}
where $\alpha$, $\beta$ are the weight balancing parameters.


\section{Experiments and Results}\label{experiments}

\subsection{Experiment on Traditional 3D Matting Methods}\label{Exp_Trad_3D_Matting}
In order to fairly compare the effectiveness of various alpha mattes and conventional binary masks, we deploy experiments on the downstream diagnosis of benign/malignant classification of pulmonary nodules, evaluating by the area under the receiver operating characteristic (AUROC). 

We take the original image, the mask processed in different ways, and the alpha mattes obtained by different methods as input, respectively, and conduct classification experiments on various types and scales of models, including Densenet 121~\cite{huang2017densely}, Resnet 34, and Resnet 50~\cite{he2016deep}. 
Each experiment is repeated four times with different seeds for robustness. 
Under the same condition, it can be assumed that the more diagnostic information available in the input, the higher the classification performance. 

\begin{figure}[!t]
	\begin{center}
		\begin{overpic}[width=\textwidth]{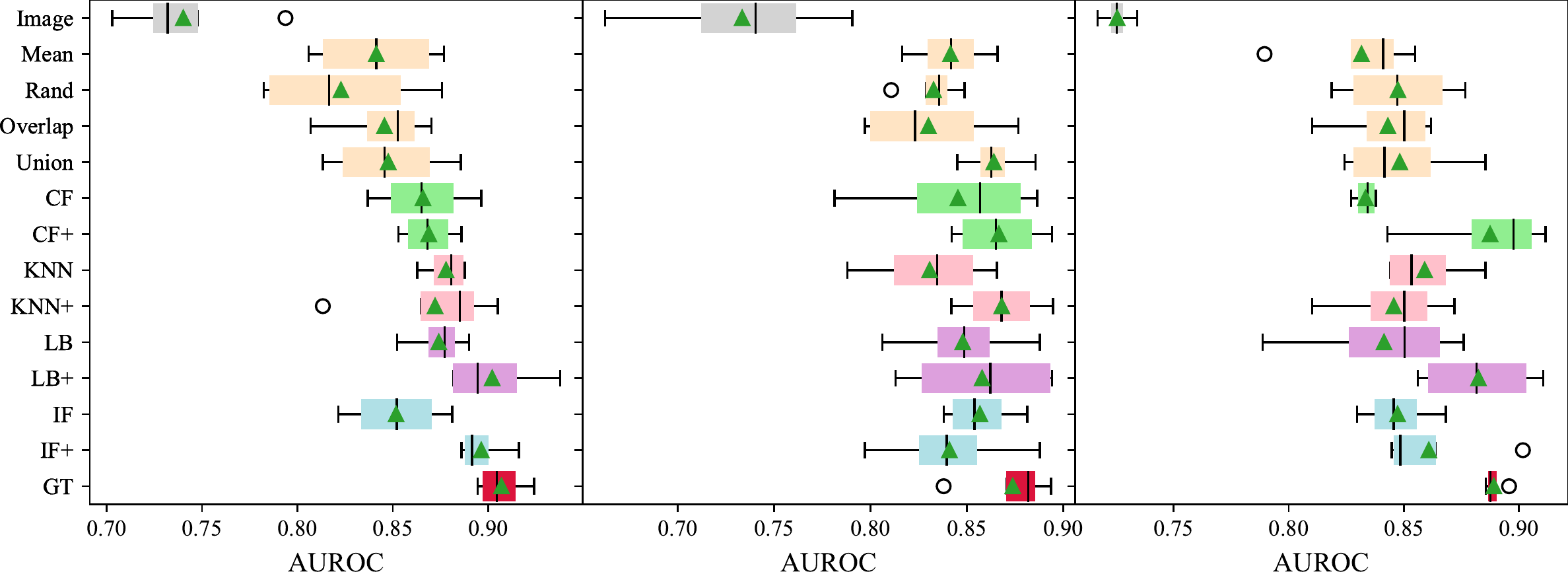}
			\put(13,-2.5){(a) Densenet 121~\cite{huang2017densely}}
			\put(46,-2.5){(b) Resnet 34~\cite{he2016deep}}	
			\put(77.5,-2.5){(c) Resnet 50}
		\end{overpic}
	\end{center}
	\caption{
		Quantitative comparisons of the diagnosis performance between binary masks derivatives and alpha mattes as input on three deep models, i.e., Densenet 121, Resnet 34, and Resnet 50. The results are shown in (a)-(c), respectively. 
		Mean, Rand, Overlap, and Union refers to the mean, randomly selected, overlap, and union of the manual labelled binary masks, respectively.
		The names with the \textit{+} suffix stand for the corresponding optimized matting methods.
		The GT alpha matte obtains the best performance, and the optimized methods outperform their counterparts in general. Moreover, the matting methods get better results than the binary mask derivatives due to the better ability to depict lesions.
		}
	\label{fig:diagnosis}
\end{figure}

\begin{figure}[!t]
	\begin{center}
		\begin{overpic}[width=\textwidth]{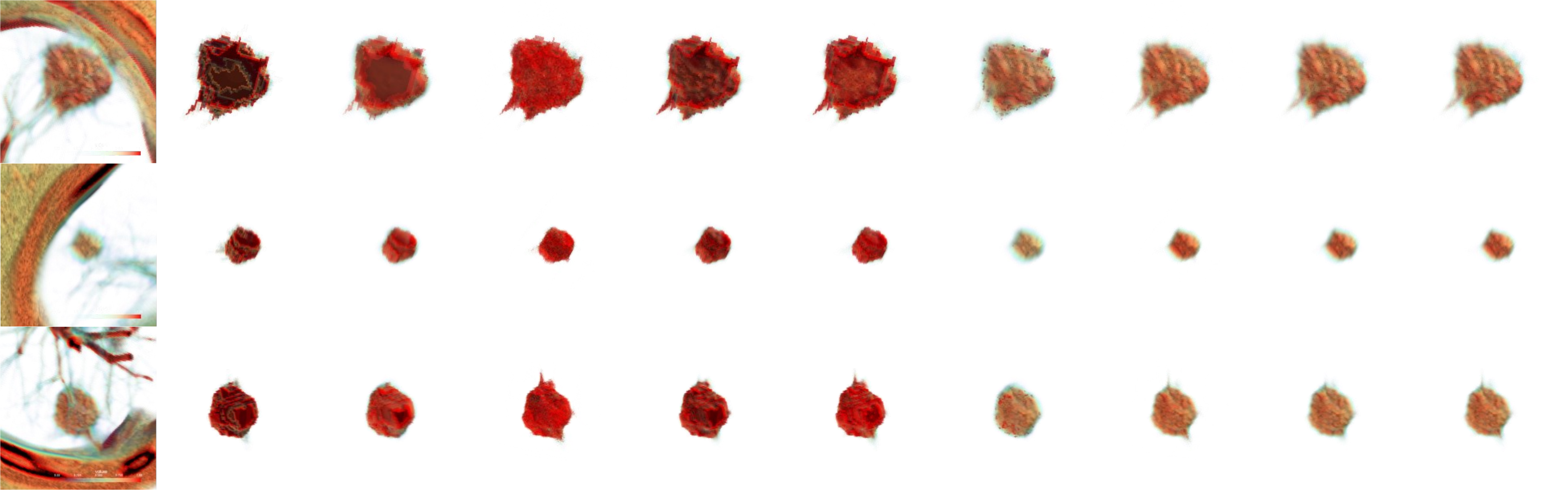}
			\put(1,-2.5){\scriptsize{(a) Image}}
			\put(10.5,-2.5){\scriptsize{(b) mMask}}	
			\put(22,-2.5){\scriptsize{(c) CF}}
			\put(31.5,-2.5){\scriptsize{(d) KNN}}
			\put(42,-2.5){\scriptsize{(e) LB}}
			\put(52.5,-2.5){\scriptsize{(f) IF}}
			\put(61.5,-2.5){\scriptsize{(g) CF+}}
			\put(70.5,-2.5){\scriptsize{(h) KNN+}}
			\put(82,-2.5){\scriptsize{(i) LB+}}
			\put(92.5,-2.5){\scriptsize{(j) IF+}}
		\end{overpic}
	\end{center}
	\caption{
		Different methods to describe the lung nodules~\cite{armato2004lung}. 
		Original 3D images and the mean masks of the labelled binary masks are shown in (a) and (b). 
		We adapt four traditional state-of-the-art 2D matting methods to 3D, shown in~(c)-(f). 
		The ability of these methods to represent the edges of the lesion is significantly improved over the binary masks, but the interior of the lesion is not numerically continuous with the edges. Hence, we modify them by calibrating the foreground to the HU value in CT images, and the corresponding results are shown in (g)-(j). The continuity of the alpha mattes has been preserved.
		The videos in supplementary files provide more details.
		}
	\label{fig:traditional_alpha_mattes}
\end{figure}
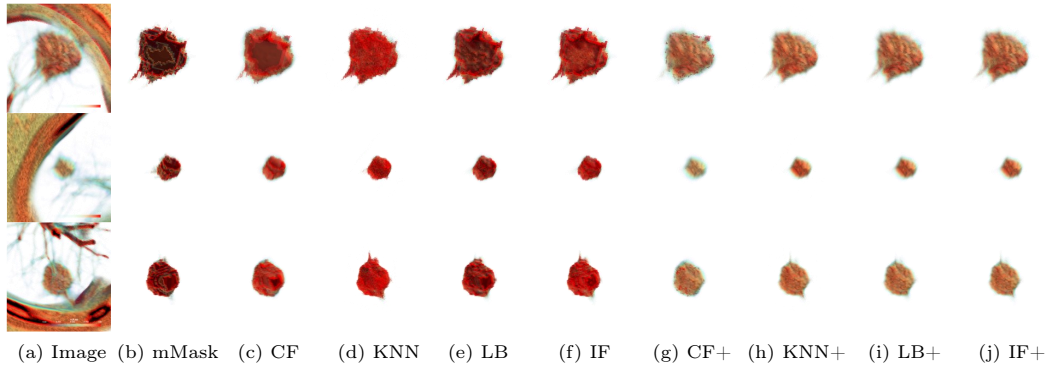

The AUROCs of the models are shown in Figure~\ref{fig:diagnosis}. 
The GT alpha matte has the best AUROC, outperforming the traditional and the optimized methods, revealing that the created dataset depicts the characteristics of the nodule lesions truly effectively. 
Also, in most cases, the alpha mattes display better performances compared with the binary mask derivatives, which illustrates that the alpha mattes are more depictable than the binary masks. 
Moreover, the optimized methods also achieve better results than their counterparts because they can better describe the informative inside of the lesions and could be a better alternative to the binary masks in lesion representation.
The original image ranks last, which may be related to its complicated input and the relatively small backbone of the classification network to capture effective features. 

Figure~\ref{fig:traditional_alpha_mattes} gives a visual display of nodules and the alpha mattes generated by the traditional and the optimized methods. 
It reveals that the alpha mattes generated by the traditional methods give a better description at the lesion's edges than the binary masks, preserving more diagnostic features such as the ground-glass shadows, but they are not continuous at the edges and interior parts of lesions, which is not consistent with the anatomical reality. 
In the optimized counterparts, the new alpha mattes have a more natural transition between the edge and interior parts of the lesions, while the internal structure information is retained. 

\subsection{Experiment on 3D Medical Matting Network}

\subsubsection{Implementation Details}
We deploy the proposed 3DMM on the annotated 3D matting dataset. 
The dataset is randomly divided into the training set, validation set, and test set, following the ratio of 7:1:2 according to the patient id. 
We augment the dataset during training by random cropping, flipping, and rotating.
The input size of the network is $96\!\times\!96\!\times\!96$.
The ADAM is the optimizer with weight decay $5\!\times\!10^{-5}$. 
The cosine annealing schedule~\cite{bochkovskiy2020yolov4} is applied after a 1-epoch long steady increasing warm-up from $0$ to the base learning rate $3\!\times\!10^{-4}$. 
The weighting parameters $\alpha$ and $\beta$ are set to $10$ due to the grid search. 
We repeat the training four times with re-splitting the dataset each time, and all the models are trained with $300$ epochs to ensure the network fully converge and batch-size $16$ from scratch using the Intel Xeon 6252 CPU and eight NVIDIA RTX 3090 graphics cards.

	
\subsubsection{Results} 
\begin{figure}[!t]
	\begin{center}
		\begin{overpic}[width=\textwidth]{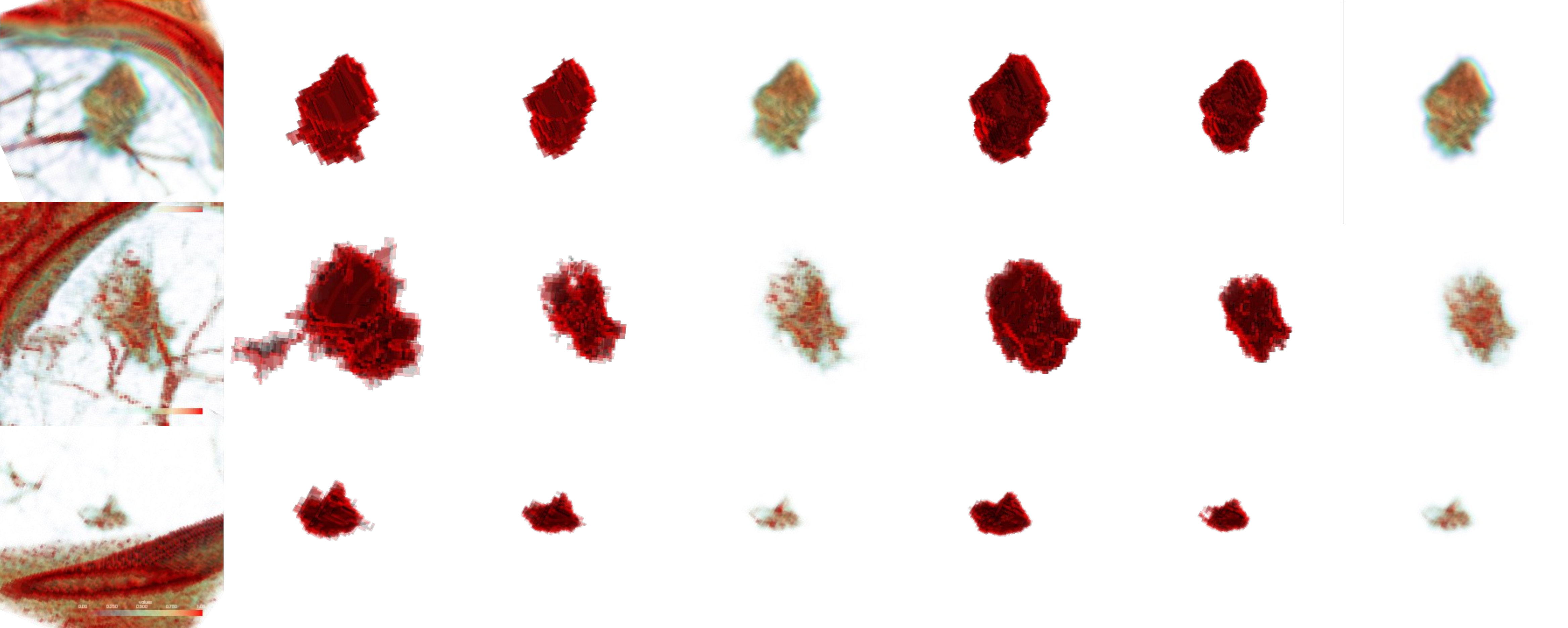}
			\put(3,-2.5){\scriptsize{(a) Image}}
			\put(15,-2.5){\scriptsize{(b) GT Union}}	
			\put(29,-2.5){\scriptsize{(c) GT Overlap}}
			\put(44.5,-2.5){\scriptsize{(d) GT Matte}}
			\put(58,-2.5){\scriptsize{(e) Pred Union}}
			\put(72,-2.5){\scriptsize{(f) Pred Overlap}}
			\put(87,-2.5){\scriptsize{(g) Pred Matte}}
		\end{overpic}
	\end{center}
	\caption{Alpha mattes and masks predicted by 3D Medical Matting Network.}
	\label{fig:dl_matting}
\end{figure}
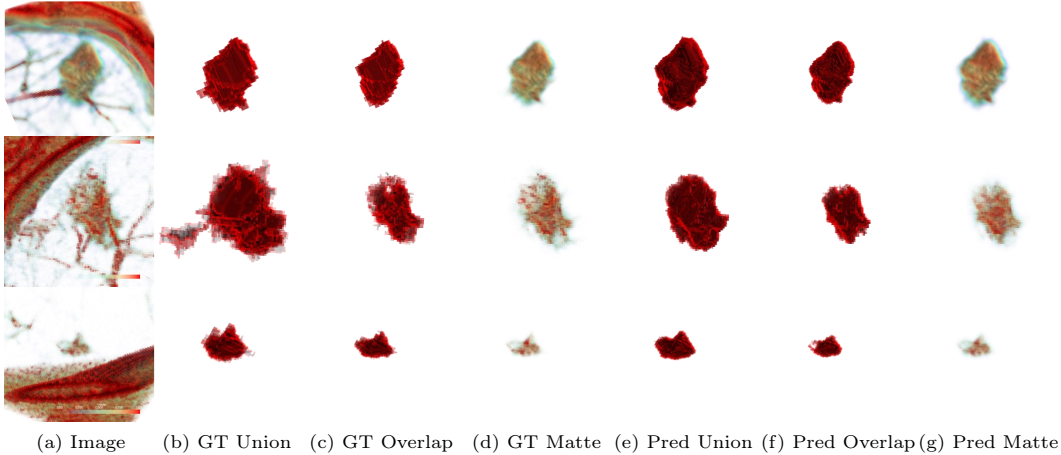

We adapt the evaluation metrics\cite{rhemann2009perceptually} widely used in image matting, namely, the sum of absolute differences (SAD), mean squared error (MSE), gradient (Grad.), and connectivity (Conn.), to 3D scenes. 
SAD and MSE are computed to evaluate the difference between the predicted and the GT alpha matte, while Grad. and Conn. focus on assessing the continuity of the predictions. 
We compare the 3DMM with the optimized methods. Moreover, we verify the role of auxiliary masks by removing them from the matting branch as an ablation study. Results are shown in Table~\ref{tab:matting_results} and Figure~\ref{fig:dl_matting}.

\begin{table}[!t]
	\caption{Quantitative comparisons with 3D medical matting algorithms${}^*$.}
	\centering
	\begin{threeparttable}
		\begin{tabular}{lc|cccc}
			\hline
			Model	&Fully Auto?	& SAD↓		& MSE↓		& Grad.↓	& Conn.↓	\\ \hline
			CF+\dag	&\usym{2717}	&152.62(±7.38)	&0.43(±0.03)	&14.96(±0.95)	&132.39(±7.10)	\\
			KNN+\dag	&\usym{2717}	&102.22(±6.68)	&0.25(±0.03)	&13.78(±1.18)	&76.01(±6.79)	\\
			LB+\dag	&\usym{2717}	&86.31(±15.52)	&\bf{0.16(±0.05)}	&\bf{6.23(±1.45)}	&\bf{66.89(±14.80)}	\\
			IF+\dag	&\usym{2717}	&\bf{79.71(±8.68)}	&0.18(±0.03)	&7.88(±0.97)	&69.09(±8.17)	\\
			\hline
			3DMM\ddag	&\usym{2713}	&106.33(±10.51)	&0.25(±0.04)	&6.75(±0.87)	&73.17(±9.24)	\\
			3DMM	&\usym{2713}	&\bf{99.42(±6.42)}	&\bf{0.24(±0.02)}	&\bf{6.37(±0.75)}	&\bf{69.25(±5.87)}	\\ \hline
		\end{tabular}
		\begin{tablenotes}
			\item[*] \scriptsize{The results are in the format of mean (± std). MSE is scaled by $1\!\times\!10^{3}$, and SAD, Grad., and Conn. are scaled by $1\!\times\!10^{-2}$. 
            The best results of the optimized traditional 3D methods and deep-learning-based methods are shown in bold, respectively.} 
			\item[\dag] \scriptsize{For fairness, the corresponding instances contributing to the GT alpha mattes are not counted in evaluation.}
            \item[\ddag] 3DMM without auxiliary masks. 
		\end{tablenotes}
	\end{threeparttable}
	\label{tab:matting_results}
\end{table}

In Table~\ref{tab:matting_results}, it is shown that the proposed 3DMM ranks second in Grad. and Conn. metrics and third in SAD and MSE.
Despite the simplification of the network, 3DMM is comparable to the optimized methods in numerical results, which reveals that the deep learning-based approach is practical to the 3D matting task. 
Moreover, the optimized methods take advantage of the prior information provided by the trimap, while 3DMM does not require additional human involvement, and it is capable of parallel inference. 

The ablation study illustrates that the proposed auxiliary mask mechanism can further improve the matting performance.
In other words, using the binary segmentation results as a guide can effectively improve the matting performance. 
This experience can enable the rapid implementation of other medical matting tasks. 
Moreover, the variance of the network with auxiliary masks is smaller than the others, which reveals that the auxiliary mask mechanism is beneficial to the models' robustness. 

From Figure~\ref{fig:dl_matting}, the multi-task network achieves good results for both masks and alpha matte prediction, and the detailed features of the lesions are better expressed. 
It is worth mentioning that because the deep-learning-based method learns potential rules from a set of data, some individual labelling errors have been corrected. The GT union mask of the second row in Figure~\ref{fig:dl_matting} gives an example.

\section{Conclusions} \label{conclusion}
This work comprehensively investigates image matting in 3D medical scenes for the first time, which illustrates that alpha mattes are more expressive in depicting the lesions than binary masks qualitatively and quantitatively. 
We adapt four state-of-the-art methods in 2D to 3D scenes and customize the methods to better characterize the information inside the lesion for medical images and give a better description at the edges than the binary masks. Based on those methods, the first 3D medical matting dataset is constructed efficiently, and downstream experiments illustrate the validity of the alpha mattes labels. The dataset will be released to the community. Meanwhile, a fully automatic matting method, 3DMM, is proposed to benchmark the study. It is the first study of the 3D deep-learning-based matting method. The trimap-free property makes it more convenient to use, and the auxiliary branches provide binary segmentation masks in different confidence levels to expand the application range.

\bibliographystyle{nips}
\bibliography{mybibliography}

\end{document}